# semanticSBML 2.0 – A Collection of Online Services for SBML Models

www.semanticsbml.org/semanticSBML

Falko Krause, Marvin Schulz, Timo Lubitz, Wolfram Liebermeister
Humboldt-Universität zu Berlin, AG Theoretische Biophysik

## Introduction

SBML (Systems Biology Markup Language) models of biochemical network models are a fast, efficient, and widely spread effort in analyzing pathways and networks. Vast amounts of software tools provide the handling of SBML files.
semanticSBML 2.0 is an advanced, improved, and extended version of the systems biology tool semanticSBML. While the latter had to be downloaded, installed and furthermore was limited in capabilities, the newly introduced and browser-enabled systems biology platform semanticSBML 2.0 can be accessed online from every operating system. It is a collection of numerous online services for the processing of SBML (Systems Biology Markup Language) files. It does not only comprise of the former semanticSBML features, but also offers many new tools for the annotation, merging, building, checking, and visualization of SBML models.
The processing of SBML models is based on the model elements and their semantic annotations. These annotations are identifiers of specific syntax linking the model elements to different web resources holding immense biochemical information (for instance KEGG, ChEBI, SBO). Our internal database *libSBAnnotation* is the annotational foundation of our SBML tools and holds identifiers from numerous web resources, also including cross-links among them.

## Features

semanticSBML 2.0 is an online collection of services for the work with biochemical network models in SBML format. After uploading one or more SBML models, the following features are available:

### 1. Diff/Merge/Split/Edit

- the 'Model Diff' feature aligns several SBML models in a tree view and detects similarities and conflicts between them; we distinguish between the whole model difference and the differences between single element attributes
- the 'Merge' feature merges two or more SBML models to one big model, taking into account equalities and similarities in the model elements; these relations are evaluated by the elements SBML identifiers or the given MIRIAM annotations

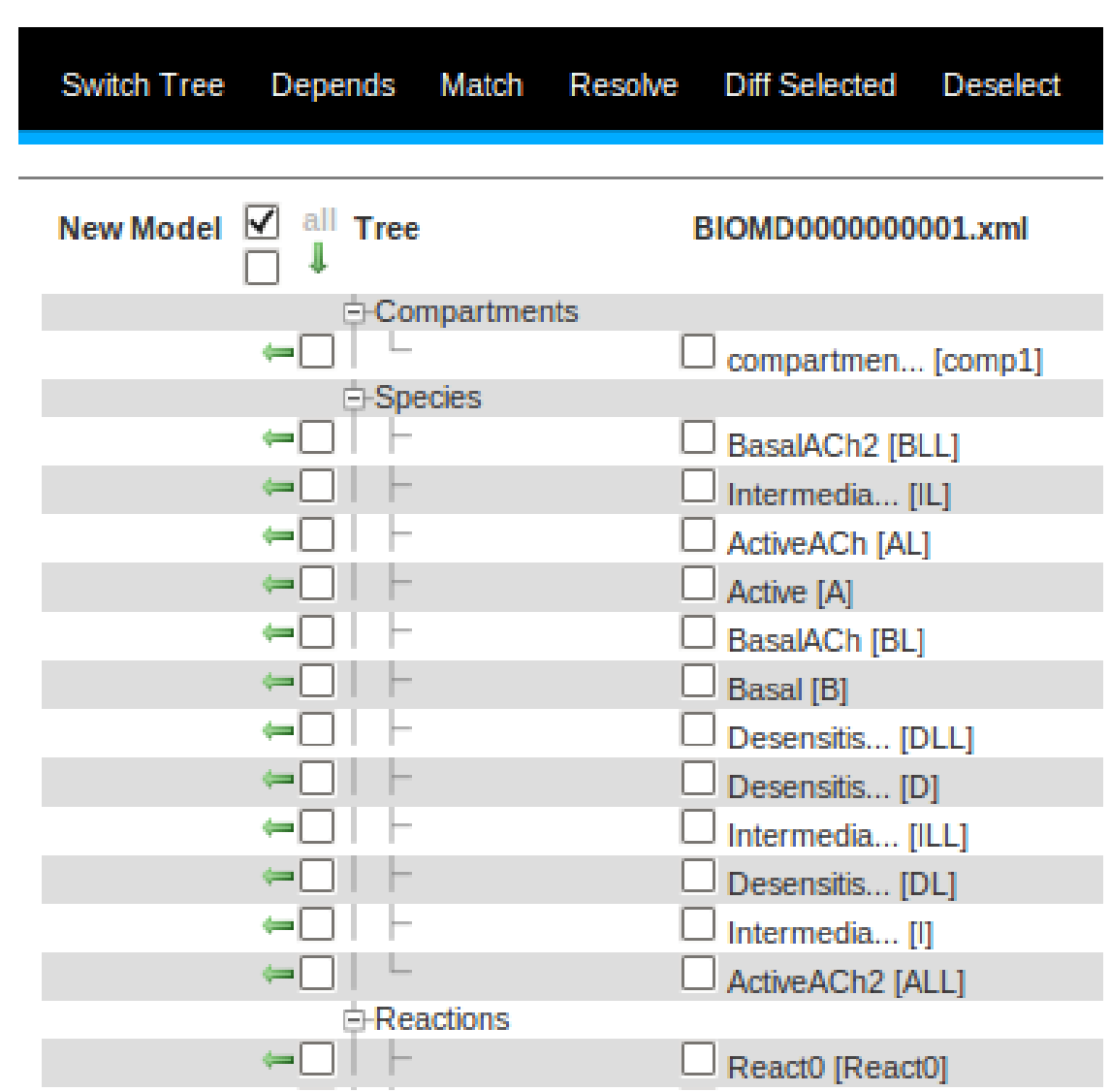

- 'Edit' provides the ability to edit the annotations of SBML models
- the 'Split' feature is able to split SBML models into smaller submodels; it is crucial to keep the mutual dependencies of the model elements in mind to construct a valid submodel

**2. Visualizator**

Generates a graphical view on the network of SBML models using GraphViz. Furthermore, using the 'SBML Wobble' feature, a flexible network graphic is created. An alternative to this visualization technique is 'SBML Browse', a feature displaying the SBML model practically in the internet browser itself.

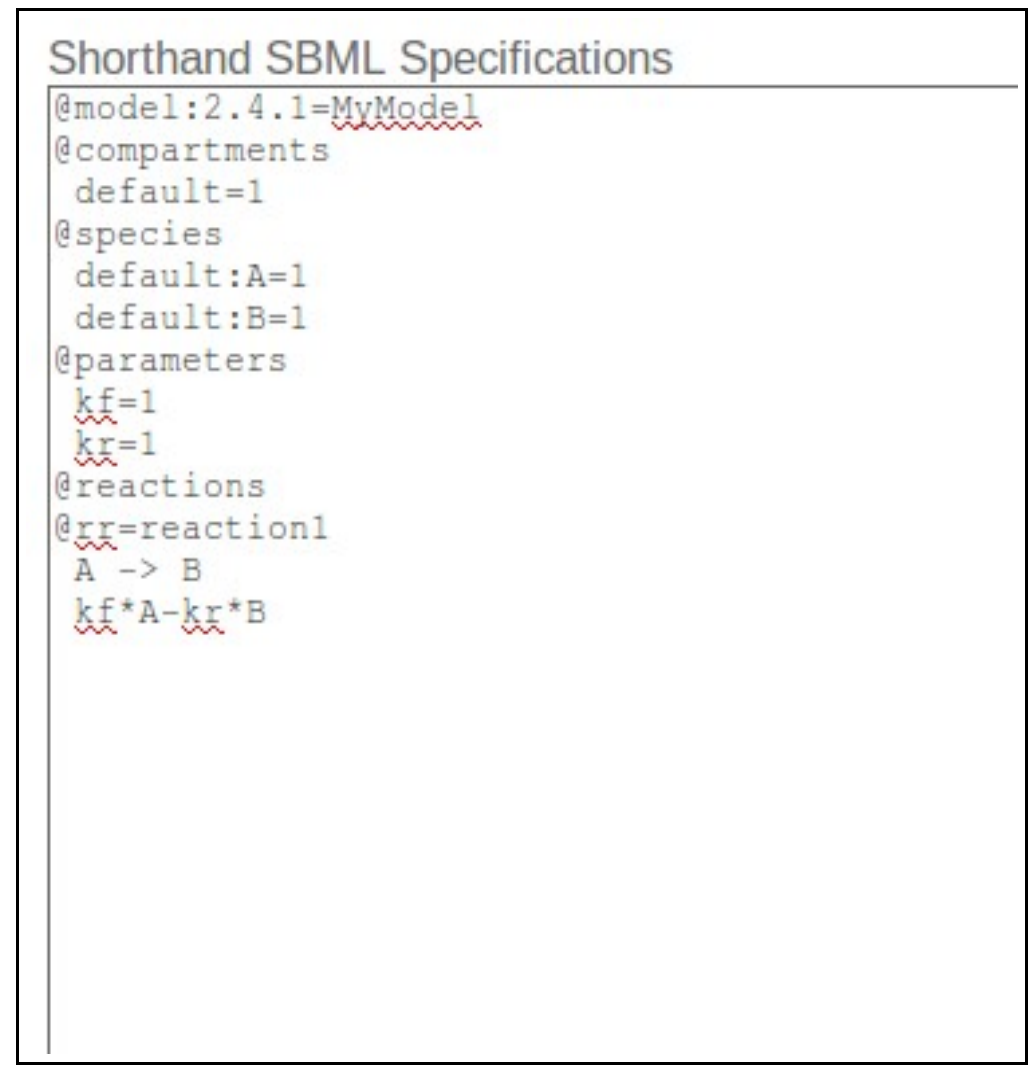

**3. Creation of SBML models**

Using the feature 'Shorthand SBML', the user can create an own SBML model. The shorthand notation is much easier to read, write, and use than the actual SBML notation, and can be translated to an SBML model using a python based tool. On our Web site we are providing an online editor for the use of shorthand SBML.

**4. RESTful API**

Many features of semanticSBML 2.0 can be accessed by a RESTful interface (Representational State Transfer), which allows the user to integrate its features in own software applications or Web sites.

**5. Automatic assignment of SBO terms**

The terms of the Systems Biology Ontology (SBO) can be automatically assigned to a models kinetic rate laws and the local parameters representing kinetic constants by just one click.

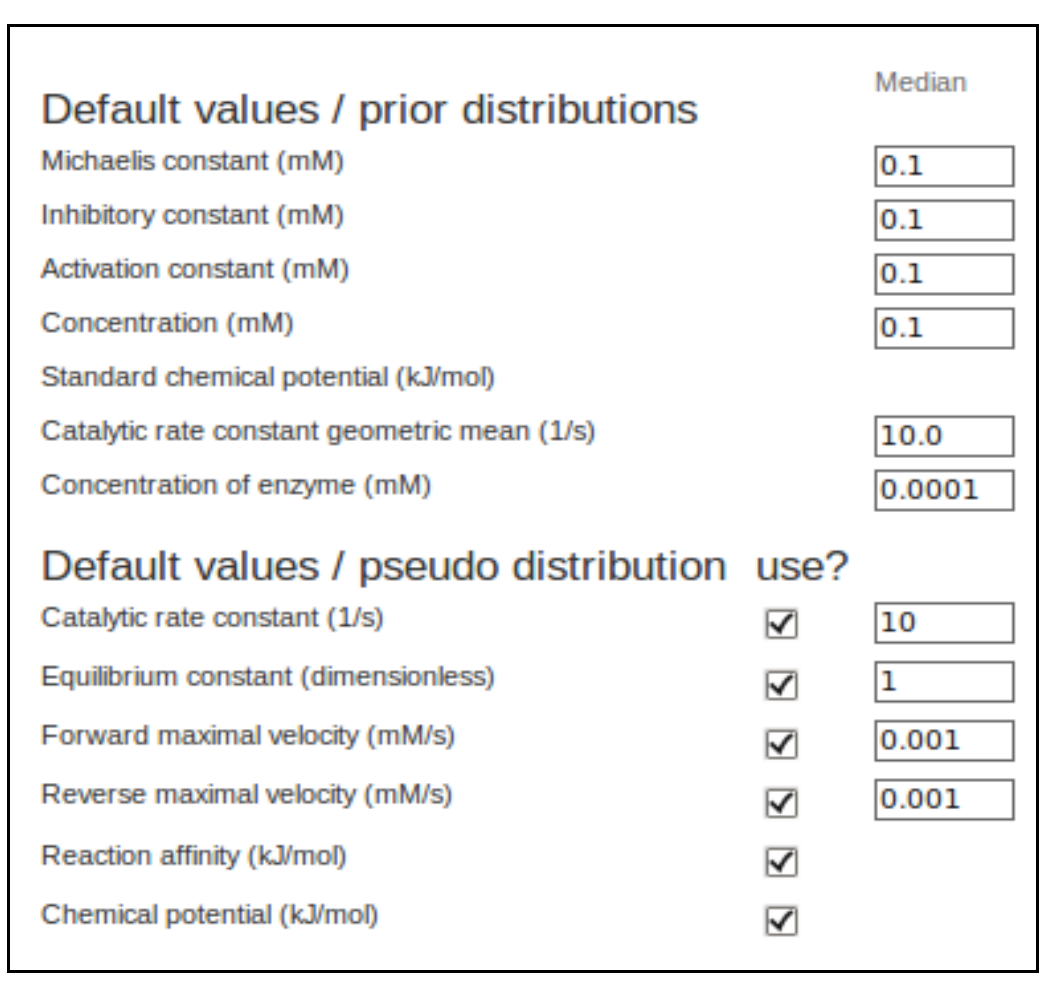

**6. Parameter balancing**

For populating an SBML model with modular rate laws, it is crucial to be provided with large sets of kinetic data to describe the rate laws. Usually, it is tedious to collect these data and often they are simply not available. 'Parameter balancing' is a tool complementing incomplete kinetic data sets by performing parameter estimates within a Bayesian framework. The produced kinetic data set is thermodynamically consistent by construction. After parameter balancing, the kinetic rate laws including the balanced kinetic data set can be inserted into the SBML model.

## 7. Clustering/similarity search

The uploaded models can be ranked and clustered by their annotation similarities and arranged in a graph view. The annotations of the SBML model elements can be removed, added, or altered by using our annotation tool 'Annotate your Model' (AYM).

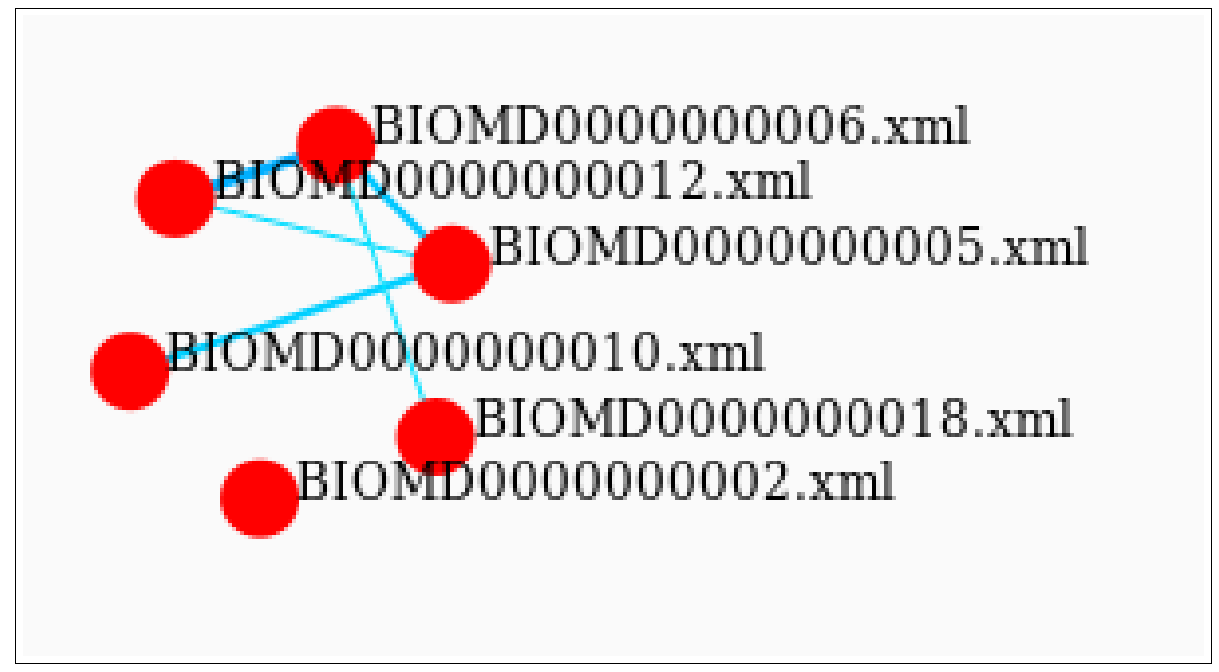

## 8. libSBAnnotation

The 'libSBAnnotation' is an internal database of biological entities and relations on which the features of semanticSBML 2.0 are based. By browsing the libSBAnnotation ontology, the user is able to discover synonyms, cross-references, and chemical relations for the elements of the uploaded SBML models. The search results are visualized in form of an editable tree view, linking to the source databases.

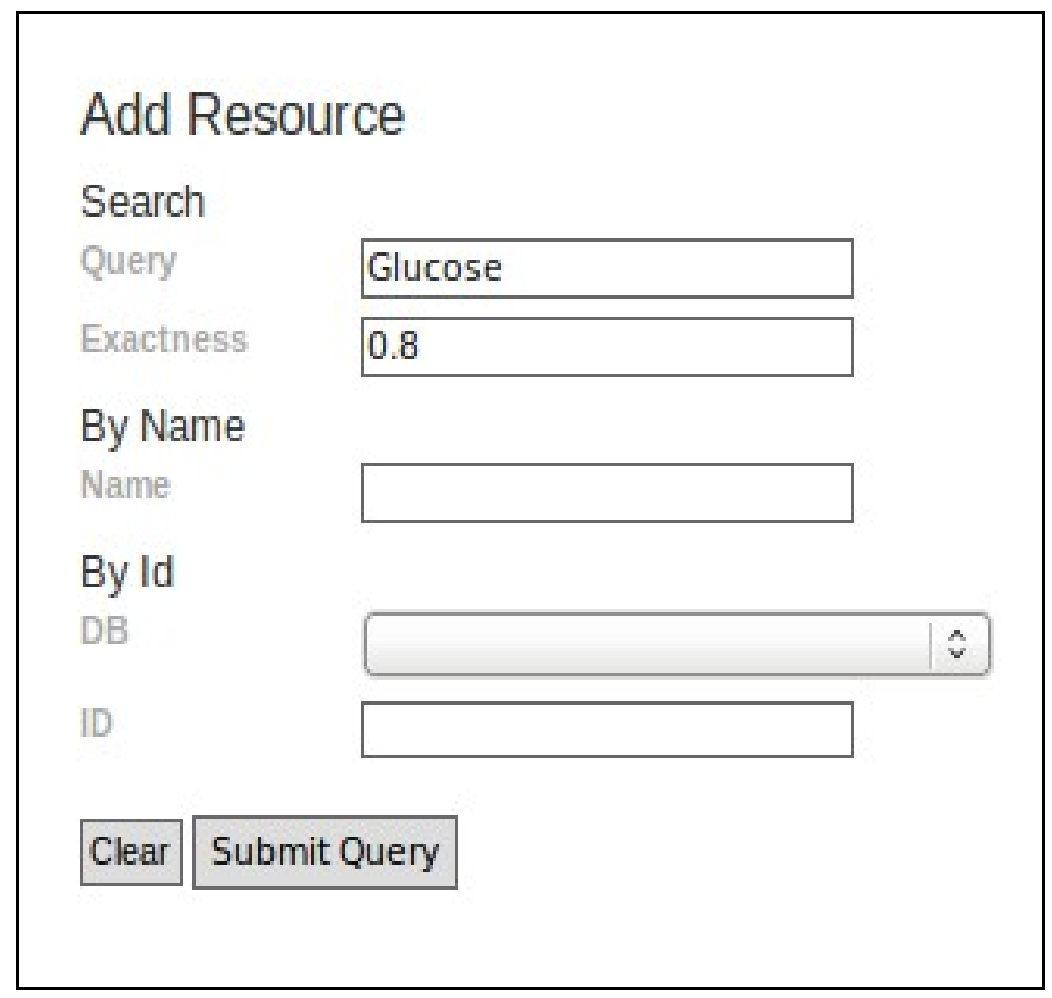

## Acknowledgements

We want to thank all scientists collaborating with us on the development of semanticSBML 2.0. Especially we want to emphasize the provided tools of Neil Swainston and Kieran Smallbone from the Manchester Institute of Systems Biology ("SBML Browse"), and Prof. Darren Wilkinson from the School of Mathematics and Statistics in Newcastle ("Shorthand SBML").